\documentclass[11pt]{article}

\usepackage{lscape,epsfig,cite,texnames,verbatim}
\usepackage{amsmath,amssymb}
\usepackage{color}
\usepackage{times}
\usepackage{float}
\floatstyle{plain}
\restylefloat{figure}

\newcommand{\grad}{\mbox{\boldmath{$\nabla$}}}
\newcommand{\iunit}{{\rm i}}
\newcommand{\lap}{\nabla^2}
\newcommand{\vdot}{\cdot}
\newcommand{\Wave}{\Psi}
\newcommand{\wave}{\psi}

\newcommand{\eps}{\varepsilon}
\newcommand{\emass}{m_{\rm e}^{}}
\newcommand{\vecs}[1]{\mathbf{#1}}
\newcommand{\brac}[1]{\tilde{#1}}

\newcommand{\eaigu}{\mbox{\'{e}}}
\newcommand{\acflex}{\mbox{\^{a}}}
\newcommand{\egrave}{\mbox{\`{e}}}
\newcommand{\agrave}{\mbox{\`{a}}}
\newcommand{\auml}{\mbox{\"{a}}}
\newcommand{\ouml}{\mbox{\"{o}}}
\newcommand{\uuml}{\mbox{\"{u}}}

\newcommand{\Mm}{\mbox{$\mathfrak{M}$} }
\newcommand{\mm}{\mbox{$\mathfrak{m}$} }
\newcommand{\Bm}{\mbox{$\cal{B}$} }
\newcommand{\Rm}{\mbox{$\cal{Q}$} }
\newcommand{\F}{\mbox{$\cal{F}$} }
\newcommand{\Y}{\mbox{$\cal{Y}$} }
\newcommand{\dY}[1]{\mbox{${\rm det}(\Y_{#1})$} }
\newcommand{\dF}[1]{\mbox{${\rm det}(\F_{#1})$} }

\newcommand{\dFe}[1]{\mbox{${\rm det}(\F {\rm e}_{#1})$} }
\newcommand{\dFo}[1]{\mbox{${\rm det}(\F {\rm o}_{#1})$} }
\newcommand{\Lag}{\mbox{$\cal{L}$} }
\def\Eg{\hbox{E.g.}\ }
\def\eg{\hbox{e.g.}\ }
\def\ie{\hbox{i.e.}\ }

\title{New Approach for the Electronic Energies of the
  Hydrogen Molecular Ion\\[0.5cm]}

\author{Tony~C.~Scott\textsuperscript{1}, 
  Monique Aubert-Fr{\eaigu}con\textsuperscript{2},
  and Johannes Grotendorst\textsuperscript{3} \\[0.5cm]
\footnotesize
    \textsuperscript{1}  Zentralinstitut f\"{u}r Angewandte Mathematik,
     Forschungszentrum J\"{u}lich GmbH, 52425 J\"{u}lich, Germany\\
\footnotesize
    Institut f\"{u}r Physikalische Chemie, RWTH Aachen, 52056 Aachen, Germany\\
\footnotesize
    Institut f\"{u}r Organische Chemie, Fachbereich Chemie,
\footnotesize
    Universit\"at Duisburg-Essen, 45117 Essen, Germany \\
\footnotesize
    email: scott@pc.rwth-aachen.de\\[0.5cm]
\footnotesize
    \textsuperscript{2}  LASIM, CNRS et Universit{\eaigu} Lyon 1,
    Campus de La Doua, B{\acflex}timent Alfred Kastler, 69100 Villeurbanne C{\eaigu}dex, France\\
\footnotesize
   email: frecon@lasim.univ-lyon1.fr\\[0.5cm]
\footnotesize
    \textsuperscript{3}  Zentralinstitut f\"{u}r Angewandte
    Mathematik, Forschungszentrum J\"{u}lich GmbH, 52425
\footnotesize
    J\"{u}lich, Germany\\
\footnotesize
email: j.grotendorst@fz-juelich.de
}

\begin{document}

\maketitle

\begin{abstract}
\par 
\noindent
Herein, we present analytical solutions for the electronic energy 
eigenvalues of the hydrogen molecular ion H$_2^+$,
namely the one-electron two-fixed-center problem.  These
are given for the homonuclear case for the countable infinity
of discrete states when the magnetic quantum number $m$ is zero
\ie{} for ${}^2 \Sigma^+$ states.
In this case, these solutions are the roots of a set of two 
coupled three-term recurrence relations.
The eigensolutions are obtained from an application of
{\em experimental mathematics} using Computer Algebra as
its principal tool and are vindicated by numerical and
algebraic demonstrations.  Finally, the mathematical
nature of the eigenenergies is identified.\\
\vspace*{2ex}
\par 
\noindent
{\bf PACS:~}
31.15.-p, 31.15.Ar, 02.70.Wz, 31.50.Bc,  31-50.Df\\

\end{abstract}

\section{Introduction}
\label{sec:intro}
Although, there are well established software packages in the
area of quantum chemistry such as GAUSSIAN\cite{gaussian},
MOLPRO\cite{molpro} and GAMESS\cite{gamess}
which allow to obtain approximate numerical solutions to
a number of fair sized molecules, the simplest
molecule namely the hydrogen molecular ion, a quantum mechanical
three-body problem, still remains mathematically intractable.

In the fixed nuclei approximation, it is well known that the
Schr\"{o}\-dinger wave equation - a second order 
partial differential equation (PDE) - of the problem of one electron 
moving in the field of two fixed nuclei can be separated
in prolate-spheroidal coordinates\cite{arfken}.  
These coordinates allow a separation of variables that results in two
non-trivial ordinary differential equations (ODE), and
hence two eigenparameters: the energy
parameter $p^2$, and a separation constant $A$ related to the total orbital
angular momentum and the Runge-Lenz vector.

We note that asymptotic expansions for 
small or large internuclear distances $R$ have been obtained.
A very comprehensive presentation of the energy eigenvalues for
the ground state and a number of exited states is shown in 
the work of \v{C}\'{\i}\v{z}ek {\em et al.}\cite{cizek}. 
These could almost constitute analytical solutions but the resulting series
are divergent though asymptotic\cite{morgan} and therefore useful
only at large internuclear distances. 
Another complication is that for the homonuclear case,
every {\em gerade} energy $E_g$  (wave function symmetric under exchange of nuclei) has
a counterpart {\em ungerade} solution (wave function antisymmetric under
exchange of nuclei) whose energy $E_u$ has exactly the same $1/R$ expansion.
This makes the calculation of exchange energy splittings 
$\Delta E = E_{u}-E_{g} $ 
very elusive to calculate at large $R$, although there are specialized
methods for recovering these splittings (\eg{} see \cite{Scott0}).

Even recently, there has been examination of series in small $R$
limited to the ground state short-range interaction energy\cite{italien} 
but we still have no further insight into 
the actual mathematical nature governing the energy eigenvalues.
We also cite the work
of Demkov {\em et al.}\cite{demkov} but their analytical solutions
correspond to a peculiar charge ratio depending on the internuclear
distance and therefore not physically useful.

Thus, complete analytical solutions of the eigenstates of
H$_2^+$, in areas of molecular interest, such as \eg{} the region 
near the equilibrium internuclear distance (bond length)
of the ground state remain elusive.

A wide variety of {\em numerical} methods have been used to solve 
the H$_2^+$ problem in this case.
For example, Bates, Ledsham and Stewart\cite{oldh2pb} used recursion
and continued fractions. Hunter and Pritchard\cite{oldh2pc} used
matrix methods and Rayleigh quotient iteration. Madsen and
Peek\cite{oldh2pd} used power series and associated Legendre expansions 
to set up two equations whose simultaneous solution then gave the two
eigenparameters.
An accurate way to obtain energies and wavefunctions for the
one-electron two-center problem is provided by the program 
ODKIL conceived by Aubert-Fr\'{e}con {\em et al.}\cite{moniquea,moniqueb}
based on a method by Killingbeck.
As of the 1980s, it was possible to calculate the 
eigenenergies and the eigenfunctions of the discrete states
of H$_2^+$ with a rapid FORTRAN program.
Yet, complete analytical solutions have so far remained elusive:
the classical $N$-body problem cannot be solved in closed form 
for $N \ge 3$ and the quantum counterpart is even worse by virtue of
being an eigenvalue problem.

The approach used here is called ``experimental mathematics'',
an unorthodox approach involving multi-disciplinary activities by 
which to find new mathematical patterns and conjectures.  The goal
in this context is to search and find mathematical structures and
patterns to be re-examined with more ``rigor'' at a later stage.
The level of rigor is of course relative: in dealing with
a difficult problem in applied Mathematics, we cannot approach
the level of rigor demanded in number theory.  Nonetheless,
we desire demonstrations sufficiently convincing to the molecular
physicist.

The present work will involve a combination of methods, results and procedures
from different areas.  We first start with results from what is
called: {\em dimensional scaling}.  It has been known for
some time that the Schr\"{o}dinger wave equation can be generalized
to an arbitrary number of dimensions $D$ which can be subsequently
treated as continuous variable\cite{dima,dimb}.  
In the limit as $D \rightarrow 1^+$, the hydrogen molecular ion
becomes the double well Dirac Delta function model which can
be solved exactly\cite{Scott1} in terms of the 
{\em Lambert W function}\cite{lamberta,lambertb}.
Dimensional scaling applied to H$_2^+$ has been studied
at length by Hershbach's 
group\cite{hershbacha,hershbachb,hershbachc,hershbachd},  in particular, by
Frantz\cite{hershbacha}, Loeser and 
Lopez-Cabrera\cite{hershbachb,hershbachc}.  The latter work
provides even more insight into the mathematical relationship between
the real H$_2^+$ at $D=3$ and its one-dimensional limit.  

Next, armed with the information provided by dimensional scaling,
we will return to the real three-dimensional formulation of 
Aubert {\em et al.}\cite{moniqueb}.  
This formulation is re-examined using a Computer Algebra System (CAS)
within the approach of experimental mathematics:
patterns and results are obtained.  The CAS used is Maple because
it is readily available to us but the results could also be
implemented on other systems.  The resulting series
expansions are verified numerically and algebraically.  In
particular, we will demonstrate that our results are independent
of choice of basis and basis size and consequently completely
general.  The end-result will be then analytically compared with the 
one-dimensional result and put on a near equal footing allowing us to 
find the mathematical category to which belong the eigenvalues of
H$_2^+$.  In view of the type of solution obtained,
a tentative ``physical'' picture is associated with the analytical
solutions.  A summary with concluding remarks is made at the end.
\addtocontents{toc}{\vspace{0.3cm}}

\section{Preliminaries - Dimensional Scaling}
\label{sec:prel}
The $D \rightarrow 1^{+} $ version of H$_2^{+}$\cite{dima,dimb} is
given by the double Dirac delta function model:
\begin{equation}
- \frac{1}{2} \frac{\partial^2 \psi}{ \partial x^2}
- q[ \delta (x) + \lambda \delta (x-R) ] \psi = E (\lambda ) \psi
\label{doublewell}
\end{equation}
where $Z_A = q $ and $Z_B = \lambda ~ q $.
The ansatz for the solution has been known since the work
of Frost\cite{frost}:
\begin{equation}
\psi ~ = ~ A e^{-d |x|} + B e^{-d |x-R|}
\end{equation}
Matching of $\psi$ at the peaks of the Dirac delta
functions positioned at $x={0,R}$ when $(\lambda =1) $ yields:
\begin{equation}
\left| 
\begin{array}{cc}
q - d & q e^{-d R} \\
q e^{-d R} & q - d 
\end{array}
\right| = 0
\end{equation}
and the energies are thus given by:
\begin{equation}
E_{\pm} = -d_{\pm}^2/2  \quad \mbox{where} \quad
d_{\pm} = q [1 \pm e^{-d_{\pm} R}]
\end{equation}
Although, the above has been known for more than half a 
century, it was not until Scott {\em et al.}\cite{Scott1} that the solution
for $d_{\pm}$ was exactly found to be:
\begin{equation}
d_{\pm} = q + W (\pm q R e^{-q R} )/R 
\end{equation}
where $\pm$ represent respectively the symmetric or {\em gerade}
solution and the anti-symmetric or {\em ungerade} solution and
$W$ is the 
Lambert W function satisfying $W(t) e^{W(t)} = t$\cite{lamberta,lambertb}.  
This function first introduced by Johann Heinrich Lambert
(1728-1777), a contemporary of Euler, 
has been ``invented'' and ``re-invented'' at various
periods in history but its 
ubiquitous nature was not fully realized within the
last decade or so.

For example, the $W$ function appears in Wien's Displacement Law
of Blackbody radiation.  In general, it has appeared in 
electrostatics, statistical mechanics, general 
relativity, radiative transfer, quantum chromodynamics,
combinatorial number theory, fuel consumption and population
growth etc\ldots (\eg{} see ref. \cite{generalapplications} and references
herein).

More recently, the Lambert $W$ function has also appeared
in ``linear'' gravity two-body problem\cite{mann} as a solution 
to the Einstein Field
equations with one spatial dimension and one time dimension $(1+1)$.
The present work also includes a generalization of the $W$ function.
Recent work\cite{martinez} shows that the $W$ function can be further
generalized to express solutions to transcendental algebraic 
equations of the form:
\begin{equation}
\exp ( \pm c ~ x) ~=~ \frac{P_N (x)}{Q_M (x)}
\label{eq:gen}
\end{equation}
where $P_N (x)$ and $Q_M (x)$ are polynomials in $x$ 
of respectively degrees $N$ and $M$ and $c$ is a constant.  
The standard
$W$ function applies for cases when $N=1$ and $M=0$ and
expresses solutions for the case of equal charges for 
eq.~(\ref{doublewell}) or equivalently the case of equal masses for the
two-body $1+1$ linear gravity problem.  The case of {\em unequal} 
charges or unequal masses corresponds to cases of higher
$N$ and $M$ values.  This form also expresses a
subset of the solutions
to the {\em three}-body linear gravity problem\cite{mann2,martinez} where one
deals with transcendental equations of the form (\ref{eq:gen})
where $M,N \rightarrow \infty $.

Some insight into the mathematical nature of the eigenenergies
of H$_2^+$ is revealed by the fact that the
eigensolutions for the electronic energies at $D \rightarrow 1^+$
and $D \rightarrow \infty $ actually bound the $D=3$ ground state
eigenenergy of H$_2^+$\cite{hershbacha,hershbachb}
as shown in Figure 1.  Moreover, the latter can 
be estimated by a linear interpolation formula\cite{hershbachc}:
\begin{equation}
E_3 (R) \approx \frac{1}{3} E_1 ({\textstyle \frac{R}{3}})~+~  
\frac{2}{3} E_{\infty} ({\textstyle \frac{2 \, R}{3}})
\end{equation}
This formula agrees with the numerically accurate eigenenergy (as
given by program ODKIL or the work of D. Frantz) to within
about 2 or 3 digits for the range of R near the bond length.
The result at $D \rightarrow \infty$ involves the extrema of
a Hamiltonian expression\cite[eq.(58)]{hershbachc}.  We
re-examined this result.
One has to consider a region of $R$ divided by
$R_c = \frac{9}{8} \sqrt{3}$.  For $R < R_c$, the
root is determined by the root of a quartic polynomial\cite{hershbachc}
and the result for $R > R_c$ is determined by a sixth degree
polynomial.  Thus, the result at $D \rightarrow \infty$ is 
algebraic.  On the other hand, the result at 
$D \rightarrow 1^+$ is in terms of an implicit special 
function, which is the Lambert $W$ function.  Given how
well this interpolation formulation works, this already
suggests what is the mathematical nature of the eigenenergies
of the true hydrogen molecular ion $(D=3)$.

We can state this in view of the work of Frantz
{\em et al.}\cite{hershbacha} who showed that the 
$D$-dimensional problem could be decoupled into two coupled
ODEs for $2 \le D < \infty$ and how a particular energy eigenvalue
for a given $D$ could exactly express the solution of 
another eigenvalue for an excited state at a dimension $D+2$ through
a precise re-scaling.

\addtocontents{toc}{\vspace{0.3cm}}
\section{Three-dimensional H$_2^+$}
\subsection{Starting Formulation}
The Schr\"{o}dinger Wave Equation for H$_2^{+}$ in atomic
units is given  by:
\begin{equation}
\left[ 
\Delta + 2 \left( \frac{Z_A}{r_A} + \frac{Z_B}{r_B} \right) + 2 E \right] 
\psi = 0
\end{equation}
As mentioned before, this is separable into prolate-spheroidal 
coordinates:
\[
\begin{array}{cclcl}
\xi & = & (r_A + r_B )/R & , & 1 \le \xi < \infty \\
\eta & = & (r_A - r_B )/R & , & -1 \le \eta \le 1 
\end{array}
\]
\[
0 \le \phi \le 2 \pi
\]
\[
\begin{array}{lcl}
\Rm_{1} & = & R (Z_A - Z_B )\\
\Rm_{2} & = & R (Z_A + Z_B )
\end{array}
\]
We can write the ansatz for the eigensolution:
\begin{equation}
\psi (\xi , \eta, \phi ) = \Lambda (\xi ) M (\eta , \phi ) =
\Lambda (\xi ) G (\eta ) e^{\pm i m \phi }
\end{equation}
which allow us to obtain two coupled ODEs:
\begin{equation}
\left[
\frac{\partial}{\partial \eta } \left( (1 - \eta^2 ) 
\frac{\partial}{\partial \eta } \right)
- \frac{m^2}{1-\eta^2} + p^2 \eta^2 - \Rm_1 ~ \eta - A \right] ~
M (\eta , \phi ) =  0 
\end{equation}
\[
\left[
\frac{\partial}{\partial \xi } \left( (\xi^2-1 ) 
\frac{\partial}{\partial \xi } \right)
- \frac{m^2}{\xi^2-1} - p^2 \xi^2 + \Rm_2 ~ \xi + A \right] ~ 
\Lambda (\xi )  =  0 
\]
where $A$ is the separation constant and the eigenenergy $E$ is
expressed as:
\begin{equation}
E_{elec} ~ = ~ -2 ~ \frac{p^2}{R^2}
\end{equation}
Note that:
\begin{eqnarray}
\lim_{R \rightarrow 0} ~ A & = &  - \ell (\ell + 1) \\
\lim_{R \rightarrow 0} ~ p^2 & = & 0
\end{eqnarray}
Although the set of quantum numbers $(n , \ell , m)$ 
- the {\em united} atom quantum numbers - can be used to 
identify the eigenstates, as is the case for \eg{} program
ODKIL, it must be emphasized that only the magnetic quantum number $m$ 
is a good quantum number (resulting from the azimuthal symmetry of 
H$_2^+$ about its internuclear axis).

We follow the treatment of Aubert {\em et al.}\cite{moniqueb} 
and consider the following basis expansion for the $\eta $ coordinate:
\begin{equation}
M(\eta , \phi ) = 
~ \sum_{k=m} f_{k}^m ~ Y_{k}^m (\eta, \phi )
\end{equation}
where $Y_k^m $ are the Spherical Harmonics.  Injection of
the above basis into the ODE governing $M$ in $ \eta$ 
leads to the creation of a symmetric matrix $\F$ whose determinant
must vanish when $p$ and $A$ satisfy the eigenvalue problem:
\begin{equation}
\left[ \F (p, A) \right] |f| = 0
\label{eq:f}
\end{equation}
where
\begin{eqnarray}
F_{i,i} ( p , A ) & = &  -i (i+1)+p^2 
\left( \frac{2 i^2 - 2m^2 + 2 i-1}{(2 i +3)(2 i -1)} \right) - A, \nonumber \\
F_{i+1,i} ( p , A )  & = & -R ~ \Rm_{1}
\left( \frac{(i + m +1)(i-m+1)}{(2 i +1)(2 i +3)} \right)^{1/2} ~, \nonumber \\
F_{i+2,i} ( p , A ) & = & \frac{p^2}{(2 i +3)}
\left( \frac{(i + m +1)(i-m+1)(i+m+2)(i-m+2)}{(2 i +1)(2 i +5)} \right)^{1/2} ~,
\nonumber
\end{eqnarray}
are the non-vanishing matrix elements of $\F$.
If $\Rm_{1} = Z_A - Z_B = 0 $ \ie{} the homonuclear case, then the
pentadiagonal matrix $\F$ divides in {\it even} and {\it odd} tridiagonal 
matrices in terms of $x$ where $x=p^2$ with no {\em explicit} dependence 
on the internuclear
distance $R$ (although this is not true for the other ODE in $\xi$).  
For the $ \xi $ coordinate, we use a basis of Hylleraas functions,
\ie{} in terms of Laguerre polynomials:
\begin{equation}
\Lambda (\xi ) =  e^{-p (\xi -1)} [2 p (\xi -1)]^{m/2}
~ \sum_{n=m/2} C_{n-(m/2)} ~ \Lag_{n-(m/2)}^{m} 
[2 p (\xi -1)]
\end{equation}
\begin{eqnarray}
\left[ \Y (p, A) \right] |C| & = & 0  \\
\left[  \Y (p, A) \right]  & = & \left[ \Rm (p, A) \right] ~+ ~p~ m^2 
\left[ \Bm (p) \right]^{-1} \nonumber 
\end{eqnarray}
where
\begin{eqnarray}
b_{i,i} ( p ) & = & 4 p +2 i +1 , \nonumber \\
b_{i,i+1} ( p ) & = & b_{i+1,i} ( p )  = 
-\left[ (i - \frac{m}{2} +1) (i + \frac{m}{2} +1) \right]^{1/2} \nonumber
\end{eqnarray}
and
\begin{eqnarray}
r_{i,i} ( p, A ) & = & (2 i +1) 
\left( \frac{R ~ \Rm_{2} }{2 p} -i -1 -2 p \right)
+ \frac{m^2}{4} + i +R ~ \Rm_{2} - p^2 + A , \nonumber \\
r_{i+1,i} ( p , A)  & = &
-\left[ (i - \frac{m}{2} +1) (i + \frac{m}{2} +1) \right]^{1/2}
\times \left( \frac{R ~ \Rm_{2}}{2 p} -i -1 \right) \nonumber \\
r_{i,i+1} ( p, A ) & = & r_{i+1,i} ( p , A) ~. \nonumber
\end{eqnarray}
When $m=0$, the matrix is tridiagonal.  For $m \ne 0$, one has
to consider the inverse of the matrix $\Bm$, which is not
a band matrix.

Of course, we realize that this choice of basis is only
one of several possible choices.  The results obtained
are valid provided the results are independent of the
size of the basis and the choice of basis.

\subsection{Recurrence Relations}
The following relations apply to the homonuclear case
and when $m=0$ in which case, the band matrices are
purely tridiagonal matrices.  These are governed by  
recurrence relations namely $(A.1)$ and $(A.2)$ of 
reference \cite{moniqueb}:
\begin{eqnarray}
{\rm det} [{\Mm}_0 ] & = & 1 \nonumber \\
{\rm det} [{\Mm}_1 ] & = & {\mm}_{1,1} \nonumber \\
{\rm det} [{\Mm}_k ] & = & {\mm}_{k,k} ~{\rm det} [{\Mm}_{k-1} ] ~- ~ {\mm}_{k-1,k} ~ {\mm }_{k,k-1} ~ 
{\rm det} [{\Mm}_{k-2} ] 
\nonumber
\end{eqnarray}
Thus for $m=0$, we have the following:
\begin{eqnarray}
\dY{0} & = & 1 \nonumber \\
\dY{1} & = & 
{\frac {R}{p}}-1-2\,p+2\,R-{p}^{2}+A \label{eq:dY} \\
\dY{k+1} & = & \left(  \left( 2\,k+1 \right)  \left( {\frac {R
}{p}}-k-1-2\,p \right) +k+2\,R-{p}^{2}+A \right) \dY{k} \nonumber \\
& & \quad -{
k}^{2} \left( {\frac {R}{p}}-k \right) ^{2} \dY{k-1} \nonumber
\end{eqnarray}
For the even $\ell$ case, we have:
\begin{eqnarray}
\dFe{0} & = & 1 \nonumber \\
\dFe{1} & = & \frac{p^2}{3} - A \nonumber \\
\dFe{k+1} & = & \left( -2\,k \left( 2\,k+1 \right) +{\frac {{p}
^{2} \left( 8\,{k}^{2} + 4\,k -1 \right) }{ \left( 4\,k-1 \right)
 \left( 4\,k+3 \right) }}-A \right) \dFe{k} \nonumber \\
& & \quad -4\,{\frac {{p}
^{4} \left( 2\,k-1 \right) ^{2}{k}^{2}~ \dFe{k-1} }{ \left( 4
\,k-1 \right) ^{2} \left( 4\,k-3 \right)  \left( 4\,k+1 \right) }}
\label{eq:dFe}
\end{eqnarray}
Defining $\dFo{i} = {\rm det} (\F (i))$ for the odd $\ell$ case, we have:
\begin{eqnarray}
\dFo{0} & = & 1 \nonumber \\
\dFo{1} & = & -2 + \frac{3 p^2}{5} - A \nonumber \\
\dFo{k+1} & = & \left( - ( 2  k+1 )  ( 2  k+2) +
{\frac {{p}^{2} \left( 2 \left( 2  k+1 \right)^{2}+1+4  k
 \right) }{ \left( 4  k+5 \right)  \left( 4  k+1 \right) }}-A \right)
\dFo{k} \nonumber  \\
& & \quad -4\,{\frac {{p}^{4}{k}^{2} \left( 2  k+1 \right) ^{
2} ~ \dFo{k-1} }{ \left( 4  k+1 \right)^{2} \left( 4 k-1
 \right)  \left( 4 k+3 \right) }} \label{eq:dFo} 
\end{eqnarray}
Note that the radial equations for the hydrogen atom are governed
by two-term recurrence relations.  Thus, it suffices to find
an eigenenergy such that the coefficient $a_{k+1}$ of the
basis of Laguerre functions is zero.  This in effect truncates
the infinite series into a polynomial and consequently closed
form solutions for the eigenstates are obtained of the hydrogen atom.  This
is not possible for H$_2^+$ which is governed by {\em three}-term 
recurrence relations no matter what the choice of basis.

The band matrices for H$_2^+$ and their determinants have been injected into a 
computer algebra system.
The determinants $\dY{i}$ and $\dF{i}$ (even or odd) for $i=1,2,3 \ldots$
are multivariate polynomial-like in $A$ and $p$.  The determinants
$\dF{i}$ are true polynomials in $A$  and $p^2$.  On the other hand,
although $\dY{i}$ is a polynomial in $A$, it has also negative
powers for $p$ and thus akin to a Laurent series (Laurent polynomial)
in $p$.

It is possible to eliminate one of the unknowns by obtaining a
{\em resultant} of the two determinants $\dY{i}$ and $\dF{i}$.  
If $a$ and $b$ are polynomials over an integral domain, where
et 2 (rational) polynomial equations in 2 unknowns $A$ and $p$.
\[
a  =  a_n \prod_{i=1}^n ( x - \alpha_i )  \quad
b =  b_m \prod_{i=1}^m ( x - \beta_i )
\]
Then
\[
{\rm resultant }(a,b,x)  =  a_n^m b_m^n \prod_{i=1}^n \prod_{j=1}^m
( \alpha_i - \beta_j )
\]
This can be computed from the Euclidean algorithm or determinant of a
{\em Sylvester} matrix and its roots will be common to those
satisfying the original set of polynomials.  
Since both expressions are true polynomials in 
$A$ only, the resultant must be in $A$. 
\Eg{} for $i=2$ (\ie{} $2 \times 2 $ matrices) 
\[
\begin{array}{l}
{\rm resultant} (\dY{i},\dFe{i} ,A) ~=~ \\
~~ \\
64/1225~p^8+512/245~p^7-256/245~(R-27)~p^6-128/245~(56~R-369)~p^5\\
~+~64/245~(2911-1037~R+27~R^2)~p^4+32/245~(13580-9405~R+948~R^2)~p^3\\
~-~32/245~(140~R^3-17780+24010~R-5571~R^2)~p^2~-~64/7~(20~R^3 -224~R^2\\
~+~481~R-161)~p-4128/7~R^3~+~19968/7~R^2~+~16~R^4~-~2880~R~+~304\\
~+~16/7~R~(28~R^3-347~R^2+791~R-252)/p~+~4~R^2~(20~R^2-108~R+85)/p^2\\
~+8~R^3~(4~R-9)/p^3~+~4~R^4/p^4
\end{array}
\]
\ie{} a Laurent polynomial in $p$ with coefficients in $R$ only.
When the size of the $i \times i$ increases, the size of the resulting
expression increases dramatically (expression swell).
However, from a numerical point of view, the most useful outcome comes
from numerically solving the simultaneous expressions for
$\dY{i}$ and $\dF{i}$ since $i$ must be sufficiently large to give
a sufficiently good result near the bond length.  
In Maple, this can be done using
the \verb+fsolve+ procedure.  To find the minimum energy 
for the ground state, it is a matter of getting derivatives
of these determinants with respect to $R$.  Combining the latter
with the condition:
\begin{equation}
\frac{\partial E_T}{ \partial R} ~ = ~ 0 \quad \mbox{where} \quad
E_T ~= ~ E_{elec} + 1/R 
\end{equation}
we get five equations in the five unknowns 
$R$, $A$, $p$, ${\textstyle \frac{\partial A}{\partial R}}$ and
${\textstyle \frac{\partial p}{\partial R}}$.  The result
has been calculated using a small Maple program.  In atomic
units, these are:
\begin{eqnarray}
 R & = & 1.997193319969992 \ldots \nonumber \\
 E_{minimum} & = & -0.6026346191065398 \ldots \nonumber
\end{eqnarray}
Note that the electronic energy, evaluated at $R=2.0$ a.u. for
comparison, is as expected exactly the reference tabulated value
of Madsen and Peek\cite{oldh2pd} \ie{} 
$ -0.6026342144949$.
An indirect way of ascertaining the accuracy of electronic energies
is to use these values in an adiabatic standard scheme to obtain
vibrational energies which are directly comparable to highly
accurate values provided by approaches that do not involve the
separability of the electronic and nuclear motions 
(\eg{} \cite{shertzer,b,c} and \cite{d,e}).
This has been done\cite{Yana} and
comparisons with values from the literature are displayed
in table \ref{tab:0}.
\begin{table}[!h]
  \centering
  \begin{minipage}{12.3cm}
 
  \label{tab:0} 
  \vspace{2ex}
  \caption{Ground State Vibrational Energies\newline Energies are in a.u., differences in cm$^{-1}$ (1 a.u. = 219474.63 cm$^{-1}$ )\newline  a) ref. \cite{shertzer},  b) ref. \cite{b}, c) ref. \cite{c},   d) ref. \cite{d},  e)  ref. \cite{e} \newline}
  \begin{tabular}{|c||c||c|c|c|}
    \hline
    & Quantum     &Present   &              &Differences \\
    System & Vibrational &Adiabatic  & Literature  & $\Delta E$ \\
    & Number $v$ & Values    & Values    &  (cm$^{-1}$) \\
    \hline \hline
    & & & & \\
    H$_2^+$ & 0 & -0.597138471 & -0.597139055$^{a)}$    & -0.13 \\
    &   &              & -0.597139063123$^{b)}$ & -0.13 \\
    & 1 & -0.587154167 & -0.587155679212$^{b)}$& -0.33 \\
    & & & & \\
    \hline
    & & & & \\
    D$_2^+$ & 0 & -0.598788594 & -0.5987876(11)$^{a)}$ &+0.22 \\
    &   &              & -0.598788784331$^{c)}$ & -0.04 \\ 
    & & & & \\
    \hline
    & & & & \\
    HD$^+$ & 0 & -0.597897521 & -0.5978979685$^{d)}$ &-0.10 \\
    &   &              & -0.5978979686$^{e)}$ & -0.10 \\ 
    & & & & \\
    \hline
  \end{tabular}
  \end{minipage}
\end{table}

In fact, given how heavy the nuclear centers are with respect to the electron,
clamping the nuclear centers is a very good approximation for the 
quantum three-body
problem represented by H$_2^+$ with the following caveat:
the approximation that the nuclei are clamped fixed in space
creates a symmetry under exchange of nuclei in the
homonuclear case.  A different picture arises when the movement
of nuclei is considered.  The mere movement of the nuclei
breaks the symmetry under exchange of nuclei and thereby leads
to a localization of the states. In this case, the work of Esry and
Sadeghpour is instructive \cite{hossein}.

However, if one stopped here, there is no pattern
from an analytical point of view.   \Eg{} setting $x=p^2$
and examining $\dFe{i}$ at low order in $A$, we have: 
\begin{description}
\item[at $i=2$:]
\[
1/35\,{p}^{2} ( -70+3\,{p}^{2} ) + ( 6-6/7\,{p}^{2}
 ) A+O \left( {A}^{2} \right) 
\]
\item[at $i=3$:]
\[
{\frac {5}{231}}\,{p}^{2} \left( 1848-126\,{p}^{2}+{p}^{4} \right) +
 ( -120+{\frac {244}{11}}\,{p}^{2}-{\frac {5}{11}}\,{p}^{4}
 ) A+O \left( {A}^{2} \right) 
\]
\item[at $i=4$:]
\[
{\frac {1}{1287}}\,{p}^{2} \left( -2162160+173316\,{p}^{2}-2772\,{p}^
{4}+7\,{p}^{6} \right)
\]
\[ 
+ ( 5040-1032\,{p}^{2} +  {\frac {6356}{195}}
\,{p}^{4}-{\frac {28}{143}}\,{p}^{6} ) A+O \left( {A}^{2} \right) 
\]
\end{description}
If we look at $A=0$ and grab the leading
coefficient $p^2$, we have the sequence\linebreak $-70, 1848, -2162160 \ldots $.
Not only are the coefficients increasing dramatically in size, 
they also alternate in sign.  Although the roots $A,p$ of these
determinants $\dY{i}$ and $\dF{i}$ converge with increasing $i$, 
the actual coefficients of these determinants and especially those of
the resultant increase in size becoming more and more
cumbersome although a CAS can handle them (up to a point). 

Moreover, we have made a particular choice of basis and the
combined set of polynomial-like expressions for the determinants
though numerically useful could be viewed more as a computational
``model'' rather than anything truly representative of the
wave function.
If we stop here, we see no pattern.  Insight comes from 
{\em inverting} the problem.

\subsection{Roots of Determinants}

The three-term recurrence relations for $\dY{i}$ or $\dF{i}$ cannot
be solved in closed form.  We start with $\dF{i}$ because it
is easier and has no explicit dependence on $R$.
Upon careful scrutiny of 
eqs.~(\ref{eq:dFe}) and (\ref{eq:dFo}), the term in $\dF{k-1}$ 
has a coefficient in $p^4$ whereas the term in $\dF{k}$ has terms
at order $p^2$.  Let us assume that $p$ is small, which is
indeed the case for small $R$.  We can therefore neglect the 
last term in $\dF{k-1}$ and the resulting two-term recurrence
relation becomes trivial to solve.  It is merely a matter 
of compounding the multiplicative terms of the recursion:
\begin{equation}
\dFe{k} ~ \approx ~ (-1)^k \prod_{j=0}^{k-1}
\left( 2\,j ( 2\,j+1 ) +A-\frac{ ( 8\, j^2 + 4\,j-1) 
~p^2 }{ ( 4\,j-1 )~( 4\,j+3 ) } \right)
\end{equation}
Solving for $A$ such that $\dFe{k} = 0$ yields:
\begin{equation}
A ~=~ -2 j ~ (2 j + 1 ) ~ + ~ 
\frac{8 j^2 + 4 j -1}{(4 j -1)(4 j +3)} ~ p^2 ~ + ~ O(p^4)
\label{eq:Aeveno}
\end{equation}
We can clearly identify the $R \rightarrow 0$ limit with
$\ell = 2 \, j$.  Similarly, for the odd case, we have:
\begin{equation}
\dFo{k} ~ \approx ~ (-1)^k \prod_{j=0}^{k-1}
\left( 2\,( j+1 )  ( 2\,j+1 ) ~ +~ A-
\frac{( 8 j^2 + 12 j + 3) p^2}{( 4\,j+5 )( 4\,j+1)} \right)
\end{equation}
\begin{equation}
A ~=~ -( 2 j +1)~(2 j + 2 ) ~ + ~ 
\frac{8 j^2 + 12 j +3}{(4 j +1)(4 j +5)} ~ p^2 ~ + ~ O(p^4)
\label{eq:Aoddo}
\end{equation}
We can clearly identify the $R \rightarrow 0$ limit with
$\ell = 2 \, j+1$.  Thus, although $\ell$ is only a
valid quantum number in the united atom limit, it is 
nonetheless feasible to use it to identify an eigenstate as 
an expansion for small $p$ (and small R).  

By the implicit function theorem, $\dF{i} =0 ~ \Rightarrow ~ A=A(p^2 )$.
Moreover, the structure of the recurrence relations for $\dF{i}$
and $\dY{i}$ namely eqs. (\ref{eq:dFe}), (\ref{eq:dFo}) and
(\ref{eq:dY}) tell us that all these quantities are $i^{th}$ degree
polynomials in $A$.  If one can find all the values of $A$
such that these determinants are zero, the latter are clearly
known by the fundamental theorem of algebra.
If $\dF{i}$ as a formal series in $x$ where $x=p^2$, we can use {\em reversion}
of power series to obtain an analytical solution.  This is 
the best possible analytical result. \Eg{}, we consider:
\begin{equation}
x ~ = ~ \cos(x) \quad \Rightarrow \quad x/\cos(x) ~=~ \lambda \quad
\mbox{where} \quad \lambda ~=~ 1 \label{eq:nonlin}
\end{equation}
\begin{eqnarray}
x~+~{\frac {1}{2}}{x}^{3}~+~{\frac {5}{24}}{x}^{5}~+~ \ldots & = & \lambda
\nonumber \\
\Rightarrow x & = & \lambda~-~{\frac {1}{2}}{\lambda}^{3}
~+~{\frac {13}{24}}{\lambda}^{5} ~+~ \ldots
\nonumber 
\end{eqnarray}
The reverted series of $x$ in terms of $\lambda$ can be obtained
in a number of ways including Lagrange's method\cite{arfken}
and represents the best possible representation of an analytical
solution to the root of eq.~(\ref{eq:nonlin}).  Formally,
the infinite series in $\lambda$ is a complete solution.  The
issue of getting numbers for \eg{} $\lambda=1$ is a matter
of a summation technique.
Solutions by reversion of power series are possible via
Maple's \verb+solve+ command.  \Eg{} inverting $\dFe{3}$ yields:
\[
1/3\,x+{\frac {2}{135}}\,{x}^{2}+O \left( {x}^{3} \right) \, , \quad
-6+{\frac {11}{21}}\,x-{\frac {94}{9261}}\,{x}^{2}+O \left( {x}^{3} \right) \, ,
\]
\[
-20+{\frac {39}{77}}\,x-{\frac {8}{1715}}\,{x}^{2}+O \left( {x}^{3} \right)
\]
where $x~=~p^2$. To first order in $x$ (or $p^2$), we recover the
solutions in eq.~(\ref{eq:Aeveno}) for respectively $\ell = 0,2,4$.
The action of inverting $\dFe{i}$ produces $i$ solutions to order $O(x^i)$.  
\Eg{} if we isolate the $\ell = 0$ solution obtained from 
$\dFe{i}$ for $i=3,4,5,6$, we obtain:
\begin{eqnarray}
&i=3:& 1/3\,x+{\frac {2}{135}}\,{x}^{2}+O \left( {x}^{3} \right) \nonumber\\
&i=4:& 1/3\,x+{\frac {2}{135}}\,{x}^{2}+{\frac {4}{8505}}\,{x}^{3}+O \left( {x}^{4} \right) \nonumber\\
&i=5:& 1/3\,x+{\frac {2}{135}}\,{x}^{2}+{\frac {4}{8505}}\,{x}^{3}-{\frac {26}{1913625}}\,{x}^{4}+O \left( {x}^{5} \right) \label{eq:Al0} \\
&i=6:& 1/3\,x+{\frac {2}{135}}\,{x}^{2}+{\frac {4}{8505}}\,{x}^{3}-{\frac {26}{1913625}}\,{x}^{4}-{\frac {92}{37889775}}\,{x}^{5}+O \left( {x}^{6} \right) \nonumber \\
&& \qquad \vdot  \quad \qquad \vdot \qquad \qquad \vdot \qquad \qquad \qquad \vdot \qquad \qquad \qquad \vdot \nonumber \\ 
&& \qquad \vdot  \quad \qquad \vdot \qquad \qquad \vdot \qquad \qquad \qquad \vdot \qquad \qquad \qquad \vdot \nonumber \\ 
&& \qquad \vdot  \quad \qquad \vdot \qquad \qquad \vdot \qquad \qquad \qquad \vdot \qquad \qquad \qquad \vdot \nonumber \\ 
&i \rightarrow \infty : & 1/3\,x+{\frac {2}{135}}\,{x}^{2}+{\frac {4}{8505}}\,{x}^{3}-{\frac {26}{1913625}}\,{x}^{4}-{\frac {92}{37889775}}\,{x}^{5}+ \ldots
\nonumber 
\end{eqnarray}
What is important to note is that the coefficients are {\em stable}!
Letting $i \rightarrow i+1$ adds a term of order $O(x^{i+1})$
to the series and yields an extra solution for $\ell = 2 \, (i+1)$.
By re-injection of this solution to within order $O(x^{i})$ into
$\dFe{i}$ with computer algebra, one can see that $\dFe{i}$ is
satisfied, term by term to within that same order.  
Conversely, the coefficients of $A$ for a particular choice 
of $\ell$ even can be obtained 
from this simple algorithm:
\begin{enumerate}
\item Select value of $j$ and $\ell ~=~ 2\, j$ and desired order $N$.
\item Set $a_0$ and $a_1$ according to eq.~(\ref{eq:Aeveno})
\begin{eqnarray}
a_0 &=& - \ell \, (\ell ~+~ 1) \nonumber \\
a_1 &=& \frac{8 j^2 + 4 j -1}{(4 j -1)(4 j +3)} ~ x ~ \nonumber
\end{eqnarray}
\item For $i ~= ~ 1$ to $(N-1)$ 
\begin{enumerate}
\item Let $A_{trial}~ =~ \sum_{k=0}^{i} a_k \, x^k + a_{i+1} \, x^{i+1}$.
Note that $a_{i+1}$ is symbolic and not yet determined.
\item Substitute $A_{trial}$ into $\dFe{ \ell + i+1}$.
\item Isolate coefficient for $x^{i+1}$.
\item Solve for $a_{i+1}$ such that this coefficient is zero.
\end{enumerate}
\end{enumerate}
A counterpart result also holds for the odd case of $\ell$ \ie{} 
for $\dFe{i}$.  
This simple algorithm allows us to yield the series solution
for $A$ for any given choice of $\ell$.
At the same time, the solution of this algorithm implies that
$\dFe{i}=0$ is formally solved.

It must be emphasized that increasing $i$ merely means
adding basis functions.  There are no singularities between the
two nuclei of H$_2^+$, and we can expect the wave function to
be not only continuous but also continuously differentiable in
that regime \ie{} we expect no surprises with the basis functions 
as $i \rightarrow \infty$.  As the estimates for $A$ and $p$ are
closer and closer to the true values of the eigenparameters, 
the magnitude of the coefficients $f_i$ of eq.~(\ref{eq:f}) 
become smaller and smaller as $i \rightarrow \infty$.  In
this limit, the basis set is a valid representation of
the true wave function.

The first 10 coefficients of the series for $A(x)$ where $x=p^2$ 
for $\ell = 0$ are:
\begin{eqnarray}
A(x) & = &
1/3\,x+{\frac {2}{135}}\,{x}^{2}+{\frac {4}{8505}}\,{x}^{3}-{\frac {26
}{1913625}}\,{x}^{4}-{\frac {92}{37889775}}\,{x}^{5} \\
& & - {\frac {513988}{
9050920003125}}\,{x}^{6}+{\frac {122264}{11636897146875}}\,{x}^{7}+{
\frac {57430742}{62315584221515625}}\,{x}^{8} \nonumber \\
& & - {\frac {26237052532}{
1566426840576238265625}}\,{x}^{9}
-{\frac {1550889714543116}{
213229853673440433908203125}}\,{x}^{10} \nonumber 
\end{eqnarray}
and our computer algebra programs allow us to generate many
more such coefficients.
The first three non-vanishing terms of the Taylor series for $A(p^2)$ have
already been published for cases of small $p$ 
consistent with small internuclear distances $R$\cite{taylora,taylorb,taylorc}.
We now claim that the present algorithm provides a
means of generating the Taylor series of $A$ in small $x$
where $x=p^2$, the result being valid as $i \rightarrow \infty $
and thus {\em independent} of the
size of the truncated basis.  
Later, we will demonstrate it to be independent
of the actual choice of basis.  However, the first test
concerns numerical vindication.

\subsection{Numerical vindication of the Series for $\bf A(p^2 )$}

To vindicate the series, we obtain data entries of 
$R$, $p$ and $A$ from program ODKIL 
and inject the data entries of $p$ into the series solutions for $A$.
We then compare the latter with the value of $A$ obtained
from ODKIL for a given state.  This is done for the ground state and a few
excited states as shown in the following tables.  
The results for the ground state \ie{} 
$1 s \, \sigma_{\rm g}^{}$ $(n=1,\ell=0,m=0)$
are shown in table \ref{tab:1a} and those
of state $ 2 s \, \sigma_{\rm g}^{}$ $ (n=2,\ell=0,m=0)$
are shown in table \ref{tab:1b} demonstrating that the
same series of $A$ for a given $\ell$ works for more than
one state.  
The results for the excited state $2 p \, \sigma_{\rm u}^{}$ 
$(n=2,\ell=1,m=0)$ vindicate the series solution for $\ell=1$.
The results for state $3 d \, \sigma_{\rm g}^{}$ $ (n=3,\ell=2,m=0)$
vindicate the series solution for $A$ for $\ell=2$.

In all cases, we can see that the series obtained for $A(p^2 )$ 
works indeed like a Taylor series, working very well for small
$p$. Beyond a certain value of $R$, the series solution
rapidly degenerates.  Nonetheless, \eg{} for the ground state,
the series solution works well near the bond length 
(around $R=2$ which is underlined) and beyond.  Degradation
of the series becomes apparent at $R=5$.
\begin{table}[!h]
 \centering
\begin{minipage}{9 cm}
\caption{Convergence of ``Taylor series'' of $\rm A(p^2 )$ \newline  $ \mbox{Ground State:}~~ 1 s \, \sigma_{\rm g}^{} \quad (n=1,\ell=0,m=0)$}
\label{tab:1a}
\vspace*{0.3cm}
\begin{tabular}{|c||c|c|c|}
\hline
&
\multicolumn{2}{c|}{ODKIL (accurate)} & series (20 terms)\\ \cline{2-4}1
$R$  & p& A & A \\
\hline \hline
0&0&0&0\\
0.5&.46569679&.729927345e-1&.7299273577e-1\\
1.0&.851993637&.249946241&.2499467374\\
1.5&1.18537488&.498858904&.4988725127\\]
\underline{2.0}&1.48501462&.811729585&.8118596153\\
2.5&1.7622992&1.19023518&1.190951531\\
3.0&2.02460685&1.64100244&1.643819599\\
4.0&2.52362419&2.79958876&2.822919217\\
5.0&3.00919486&4.37769375&4.491055954\\
10.0&5.47986646&20.1332932&25.05609231\\
20.0&10.4882244&90.0528912&-1147.477000\\
\hline
\end{tabular}

\vspace*{1cm}

\end{minipage}
\end{table}

\begin{table}[!h]
 \centering
\begin{minipage}{9 cm}
\caption{Convergence of ``Taylor series'' of $\rm A(p^2 )$ \newline  $ 
\mbox{State:}~~ 2 s \, \sigma_{\rm g}^{} \quad (n=2,\ell=0,m=0)$}

\label{tab:1b}
\vspace*{0.3cm}
\begin{tabular}{|c||c|c|c|}
\hline
 &
\multicolumn{2}{c|}{ODKIL (accurate)} & series (20 terms)\\ \cline{2-4}
$R$  & p& A & A \\
\hline \hline
0&0&0&0\\
0.5&0.241110452&0.194282436e-1&0.1942824361e-1\\
1.0&0.459850296&0.711543142e-1&0.7115431427e-1\\
2.0&0.849546791&0.248466171&0.2484661714\\
3.0&1.19791141&0.510154273&0.5101542740\\
4.0&1.51924947&0.8535318&0.8535318053\\
5.0&1.82176362&1.28400188&1.284001886\\
10&3.19930169&5.12935962&5.127249696\\
15&4.51129751&12.4337232&-17315.20146\\
\hline
\end{tabular}
\end{minipage}
\end{table}

The question arises as to whether or not the series coefficients
of $A(p^2 )$ follow a pattern.  We have found none so far.  The
pattern of the changing signs $+,-$ is not one of alternating
series and thus this function is unlike all the special
functions known in the literature (such as \eg{} \cite{AbrSte70a}). 
\begin{table}[!h]
 \centering
\begin{minipage}{8.5 cm}

\caption{Convergence of ``Taylor series'' of $\rm A(p^2 )$ \newline  $ 
\mbox{State:}~~ 2 p \, \sigma_{\rm u}^{} \quad (n=2,\ell=1,m=0) $}

\label{tab:1c}
\vspace*{0.3cm}
\begin{tabular}{|c||c|c|c|}
\hline
 &
\multicolumn{2}{c|}{ODKIL (accurate)} & series (20 terms)\\ \cline{2-4}
$R$  & p& A & A \\
\hline \hline
0&0&-2&-2\\
0.5&.254186316&-1.96120498&-1.961204981\\
1.0&.53141962&-1.83001042&-1.830010419\\
2.0&1.15545177&-1.18688939&-1.186889387\\
4.0&2.35889913&1.53846448&1.538464473\\
6.0&3.43970785&5.92793017&5.927930398\\
8.0&4.4671459&12.0646853&12.07439611\\
9.0&4.97308004&15.8356448&16.60977070\\
10&5.476774&20.0920989&58.89905749\\
20&10.4882239&90.0528776&.7649129703e13\\
\hline
\end{tabular}
\vspace*{1cm}

\end{minipage}
\end{table}

\begin{table}[!h]
 \centering
\begin{minipage}{9 cm}
\caption{Convergence of ``Taylor series'' of $\rm A(p^2 )$ \newline
$\mbox{State:}~~ 3 d \, \sigma_{\rm g}^{} \quad (n=3,\ell=2,m=0)$}
\label{tab:1d}
\vspace*{0.3cm}
\begin{tabular}{|c||c|c|c|}
\hline
 &
\multicolumn{2}{c|}{ODKIL (accurate)} & series (20 terms)\\ \cline{2-4}
$R$  & p& A & A \\
\hline \hline
0&-&-6&-6.\\
0.5&0.166934253&-5.98541087&-5.985410869\\
1.0&0.335547827&-5.94115241&-5.941152409\\
2.0&0.686698811&-5.75530105&-5.755301048\\
4.0&1.51188304&-4.86085811&-4.860858108\\
6.0&2.37168861&-3.43229937&-3.432299419\\
8.0&3.09069127&-2.07684281&-2.076688125\\
10&3.69538523&-0.874720469&2.071237971\\
20&6.12806789&7.31365225&5232651466.\\
\hline
\end{tabular}
\end{minipage}
\end{table}
Nonetheless, there is something of a pattern for a given
series when modifying the quantum number $\ell$, term by term.
The first two terms $a_0$ and $a_1$ follow a pattern in $\ell$
according to \eg{} (\ref{eq:Aeveno}) for even $\ell$. No such
simple pattern exists for the next term $a_2$.  However,
if one solves for $a_2$ in terms of $a_0$ and $a_1$ for
a high value of $\ell$, say $\ell=\ell_{max}$, one obtains
a polynomial formula for $a_2$.  If one then substitutes
the general formulae in $\ell$ for $a_0$ and $a_1$ into
this polynomial expression for $a_2$: it will correctly
generate the coefficients $a_2$ not only for $\ell_{max}$
but for all $\ell = 0,1,2 \ldots \ell_{max}$. At some point,
the resulting formula will break down for a value of 
$\ell > \ell_{max}$.  This ``triangular'' relationship,
- useful because one often does calculations within for a limited
range of $\ell$ - indicates that:
\[
a_2 ~ \approx ~ \lim_{\ell_{max} \rightarrow \infty} \frac{P_{\ell_{max}} (\ell )}{Q_{\ell_{max}} (\ell )} \, ,
\]
which places us beyond eq.~(\ref{eq:Aeveno}) (or (\ref{eq:Aoddo}))
which determine $a_0$ and $a_1$ only.
However, this is subject of further exploration elsewhere.

The range of the series solution can be considerably 
improved by modifying the recurrence relation for $\dFe{}$ like so:
\begin{eqnarray}
\dFe{0} & = & 1 \\
\dFe{1} & = & \frac{y}{3} - A \\
\dFe{k+1} & = & \left( -2\,k \left( 2\,k+1 \right) +{\frac {y
\left( 8\,{k}^{2}-1+4\,k \right) }{ \left( 4\,k-1 \right)
 \left( 4\,k+3 \right) }}-A \right) \dFe{k} \\
& & \qquad -4\,{\frac {x~y
\left( 2\,k-1 \right) ^{2}{k}^{2}~ }{ \left( 4
\,k-1 \right) ^{2} \left( 4\,k-3 \right)  \left( 4\,k+1 \right) }} 
\dFe{k-1}
\nonumber
\end{eqnarray}
where it is understood $x=y=p^2$ but it is only $x$ which is treated
as a perturbation. This is simply a different representation denoted
$A=A(x,y)$ but which represents the same function $A(p^2 )$. Modifying
slightly our previous algorithm, we obtain \eg{} a modified series
solution for $A(x,y)$ for $\ell=0$:
\begin{equation}
A = \frac{y}{3}-{\frac {14}{15}}\,\frac {y~x}{(2\,y - 63)}+
\frac {14}{375}\,\frac{y^2 ~ P_1 (y) ~ x^2}{( 2\,y -63)^{3} ( 2\,y-231)}
\end{equation}
\[
\qquad
-\frac{28}{121875}\,
\frac{y^3 ~ P_3 (y)~  x^3}{ ( 2\,y - 63 )^5~ (2 \,y-231 )^2 ( 2\,y-495 ) } +
\ldots
\]
where the polynomials $P_k (y)$ of order $k$ are given by:
\begin{eqnarray}
P_1 (y) & = & 94\,y-44121  \nonumber \\
P_3 (y) & = & 166376\,y^3 + 16398492\,y^2+131745081006\,y- 13685763372435 
\nonumber
\end{eqnarray}
Note that if we inject $y=x$ into the above and make a Taylor series expansion
in $x$, we simply recover the series solution in $x=p^2$ obtained
in eq.~(\ref{eq:Al0}) for $\ell=0$.
Since the radius of convergence is determined by the closest
singularity or branch point in the complex plane, we have
\[
2 \, y - 63 = (2~p^2 ~- ~63)=0 \Rightarrow p \approx 5.6 
\]
We note that the sequence of numbers 63, 231, 495, 855,\ldots
which appear in the denominator have a pattern which can be found
using the \verb+gfun+ package\cite{gfun}.  This demonstrates
that these numbers fit a holonomic function and it is found
that these fit the pattern:
\begin{equation}
3 \, (4 \, j + 3) ( 4 \, j + 7 )
\end{equation}
We recognize it as one of the terms which appear in 
the recursion relations for $\dFe{k}$ \linebreak \ie{}
$(4 k -1)(4 k +3)$ with $k=j+1$.  However, no pattern has
(so far) been found for the polynomials $P_k (y)$.  Nonetheless,
our computer algebra routines allow us to generate this 
series to relatively high order.

Next, the sum can be calculated using non-linear transformations known as 
the Levin or Sidi transformations.
The latter involves a series transformation by
which one can accelerate the convergence of a series and even sum
divergent series (\eg{} see the work of \cite{levin,weniger}).
We take the point of view that a Taylor or asymptotic series
has all the desired ``information'', getting numbers from the
series is a matter of a summation technique. These transformations
are available in the Maple system as \verb+NonlinearTransformations+.

The best results for the ground state are obtained by applying a 
Sidi $d$ transformation
in $x$ compounded with $y$ as shown in table \ref{tab:1e}.  Even
when the modified series behaves badly, the result from the 
Sidi $d$ transformation provides reliable numbers. The
results hold up remarkably well all the way up to $R=10$ and
beyond.  Beyond $R=10$, the asymptotic series expansions as
\eg{} listed by \v{C}\'{\i}\v{z}ek {\em et al.}\cite{cizek} are more
useful.  What is important in our case, is that our series
solution works so well around the bond length and the intermediate
regime.  
\begin{table}[!h]
 \centering
\begin{minipage}{10.5 cm}

\caption{Convergence of Series A(x,y) \newline $
\mbox{Ground State Revisited:}~~ 1 s \, \sigma_{\rm g}^{} \quad (n=1,\ell=0,m=0) $}
\vspace*{0.3cm}
\label{tab:1e}
\begin{tabular}{|c||c|c|c|}
\hline
 &
\multicolumn{3}{c|}{A} \\ \cline{2-4}
$R$  & series (12 terms) & ODKIL (accurate) & Sidi-d \\
\hline \hline
 1&0.24994624090& 0.2499462409& 0.2499462410\\
\underline{2}&0.81172958404&0.8117295840&0.8117295850\\
 3&1.6410024366& 1.6410024369& 1.6410024370\\
 4&2.7995666114& 2.7995887586& 2.7995887590\\
 5&3.9638237398& 4.3776938960& 4.3776937530\\
 6&-4.6313683166e+03& 6.4536051398& 6.4536037430\\
 8&-3.7137673759e+12& 12.2262006172& 12.2261746150\\
10&-1.5608159299e+33& 20.1339450995& 20.1332931780\\
15&1.0054600411e+15& 48.8656127918&48.8223535290\\
\hline
\end{tabular}
\end{minipage}
\end{table}

\subsection{Solution for $\bf A(R,p)$}
Although we have eliminated one of the unknowns \ie{} found
$A(p^2 )$ such that the determinantal conditions for $\dF{i}$ 
are satisfied, there is still the remaining determinant 
$\dY{i}$ to address.  The recurrence relations for $\dY{i}$ 
of eq.~(\ref{eq:dY}) depend on the internuclear distance $R$
and have more structure than those of 
$\dFe{i}$ of eq.~(\ref{eq:dFe}) or $\dFo{i}$ of eq.~(\ref{eq:dFo}).
Nonetheless, we proceed in parallel to what we did for $A(p^2 )$.

To start with, we ignore the term $\dY{k-1}$ and solve the
resulting two-term recurrence relation since
all linear recurrence relations of this type
are solvable in terms of the roots of the characteristic 
polynomial obtained by assuming a $\dY{i} = f^i$ and 
then solving for $f$:
\begin{equation}
\dY{i} ~=~ 
(-2)^k~\frac{\Gamma(({2~k~p+Y_{+} +X}/{2 p})) 
~\Gamma((2~k~p + Y_{+} -X)/(2 p))}{\Gamma(-(Y_{-}-X)/(2 p))~
\Gamma(-(Y_{-}+X)/(2 p))}
\end{equation}
where
\begin{eqnarray}
X &= & \sqrt{2~p^4-p^2+R^2+2~A~p^2} \nonumber \\
Y_{+} &= & 2~p^2 ~+~ p ~-~R  \nonumber \\
Y_{-} &= & 2~p^2 ~-~ p ~+~R  \nonumber 
\end{eqnarray}
and $\Gamma$ is the Gamma function\cite{AbrSte70a}. 
This result bears some resemblance 
with the outcome of solving the eigenvalue problem for
the hydrogen atom.  In this case, solutions to the ODE
for the radial equation in the radius $r$ 
can be expressed in terms of hypergeometric functions.  
Matching the asymptotic solution at $r \rightarrow \infty$ 
with the regular solution at $r \rightarrow 0$ necessitates
the elimination of the irregular solution by forcing one 
of its coefficients - also expressed in terms of the Gamma function~-
to be zero (\eg{} see \cite{hydrogen}).  
In our case (as in the case of the hydrogen
atom), it is a matter of ensuring that the arguments for
one (or both) of the Gamma functions in the denominator of the
expression above to be $-\j$ where $\j=0,1,2 \ldots$
Thus, solving for $A$, we find that:
\begin{equation}
A(R,p) ~\approx ~ p^2~+~2\,(1+2\,\j )\, p~+~1-2 \, R~+~
2\,\j ~+~2 \,\j^2-\frac{R \,(1+2 \, \j ) }{p} \, .
\end{equation}
What remains is the identification of $j$.
Next we treat term $\dY{k-1}$ as a perturbation formally by multiplying
it by $\lambda$
with the understanding that $(\lambda =1)$.
For $ \j = 0$, the series solution for $A(R,p)$ is:
\begin{eqnarray}
A(R,p) & = & (p+1)^2-2~R-\frac{R}{p} 
~  - ~\frac{ ( p-R)^2 \lambda }{2~ p ~ ( 2\,p^{2} + 2~p -R ) } \label{eq:ARp}\\
&  & - \frac{(p-R)^2 ~P_4 (R,p) \lambda^2}{(8~p~(2~p+2~p^2 -R)^3~(3~p+2~p^2 -R ))}
\nonumber \\
&  & - \frac{(p-R)^2 ~P_{10} (R,p) \lambda^3}{(16~p~(2p+2p^2-R)^5~(3p+2p^2 -R)^2~(4p+2p^2 - R))} \nonumber 
\end{eqnarray}
where
\begin{eqnarray}
P_4 (R,p) & = & 
14\,p^4 + (13 -12 R ) \, p^3 ~+~ R \, ( 2\,R-17 ) \, p^2 + 7\,p R ^2 - R^3 \nonumber \\
P_{10} (R,p) & = & 584\,p^{10} ~- ~8\, ( 116\,R -233) \, p^9+2\, ( 256\,{
R}^{2}+969-1878\,R ) ~ p^8
\nonumber \\
& & - ~4\, ( 28\,R^3-722\,R^2 +1184\,R-165 ) ~ p^7
\nonumber \\
&& + ~R ~ ( 8\,R^3+4678\,R-1056\,R^2-1897 ) ~ p^6
\nonumber \\ 
&& +2\,R^2 ~ ( 92\,R^{2}-1200\,R ~+ ~ 1145 ) ~ p^5
\nonumber \\
&& - ~ R^3 ~ ( 12\,{R}^{2}-678\,R+1513 ) ~ p^4
- ~2\,R^4 ( 50\,R - 297 ) ~ p^3
\nonumber \\
&& + ~R^5 ~
( 6\,R-139) \, p^2 ~+~18 \, R^6 \, p ~-~R^{7}
\nonumber 
\end{eqnarray}
The series looks complicated and the presence of singularities at
every $-R + i \, p +2~p^2 ~=~ 0$ for $i=2,3,\ldots$ already tell
us that this function is unlike most special functions
in the literature.  However, the series gives very good results
as shown in table \ref{tab:2} with only 4 terms.  It does not need 
any convergence acceleration summation methods at large $R$.
The results of Table \ref{tab:2b} for 
state $ 2 s \, \sigma_{\rm g}^{}$ $ (n=2,\ell=0,m=0)$
show us that $A(R,p)$ works well for large $R$ but diverges
for small $R$.  Also shown in the table are the results
of the Sidi $d$ transformation which considerably improves the 
series solution for small $R$.

What remains is to identify the meaning of the number $j$.  By
checking the solution for excited states, we find out empirically
that:
\begin{equation}
\j ~=~ n ~-~ \ell~ -~ 1
\label{eq:j}
\end{equation}
where $n$ is the united atom quantum number.
This number $\j$ is a valid quantum number for the {\em separated}
atom limit\cite[eq.24,p.666]{oldh2pe}.
Thus, just as we match the outward and inward radial solutions
for the radial ODE for the hydrogen atom by which to determine
the eigenvalue, the eigensolution for H$_2^+$ results from
matching $A(p^2 )$ governed by the united atom quantum number $\ell$
with $A(R,p)$ governed by the separated atom quantum number 
$\j = n - \ell - 1 $.
\begin{table}[!h]
 \centering
\begin{minipage}{9.6 cm}

\caption{Convergence of $\rm A(R,p)$ \newline 
$ \mbox{Ground State:}~~ 1 s \, \sigma_{\rm g}^{} \quad (n=1,\ell=0,m=0) $ }
\vspace*{0.3cm}
\label{tab:2}
\begin{tabular}{|c||c|c|c|}
\hline
 &
\multicolumn{2}{c|}{ODKIL (accurate)} & series (4 terms)\\ \cline{2-4}
$R$  & p& A & A \\
\hline \hline
0.5&0.46569679&0.729927345e-1&0.7299778055e-1\\
1.0&0.851993637&0.249946241&0.2499480309\\
\underline{2.0}&1.48501462&0.811729585&0.8117297560\\
5.0&3.00919486&4.37769375&4.377693772\\
10&5.47986646&20.1332932&20.13329314\\
20&10.4882244&90.0528912&90.05289034\\
30&15.4919739&210.034597&210.0345960\\
40&20.4939187&380.025707&380.0257060\\
50&25.4951064&600.020452&600.0204512\\
\hline
\end{tabular}
\end{minipage}
\end{table}

\begin{table}[!h]
 \centering
\begin{minipage}{13 cm}
\caption{Convergence of $\rm A(R,p)$ \newline
$ \mbox{State:}~~ 2 s \, \sigma_{\rm g}^{} \quad (n=2,\ell=0,m=0) $}
\vspace*{0.3cm}
\label{tab:2b}
\begin{tabular}{|c||c|c||c||c|}
\hline
 &
\multicolumn{2}{c|}{ODKIL (accurate)} & series (5 terms) & Sidi-d \\ \cline{2-5}
$R$  & p& A & A & A \\
\hline \hline
0&0&0&-&-\\
0.5&0.241110452& 0.0194282436&-7.5860659805e+02&0.0211395500\\
1.0&0.459850296& 0.0711543142&-2.1405738045& 0.0718499907\\
2.0&0.849546791& 0.2484661710&0.23871213039& 0.2485999772\\
3.0&1.197911410& 0.5101542730&0.50971077859& 0.5101743643\\
4.0&1.519249470& 0.853531800& 0.85348392428& 0.8535343888\\
5.0&1.821763620& 1.28400188&1.2839939077& 1.2840020996\\
10.&3.199301690& 5.12935962&5.1293596329& 5.1293596444\\
15.&4.511297510& 12.4337232&12.433723259& 12.4337232589\\
20.&5.805158110& 23.1467952&23.146795143& 23.1467951431\\
30.&8.359177000& 54.1918175&54.191817437& 54.1918174372\\
40.&10.8899708000& 97.83692290 & 97.836923003& 97.8369230031\\
\hline
\end{tabular}

\end{minipage}
\end{table}
As suggested by Table \ref{tab:2b} the series behaves well for large $p$,
it is found that $A(R,p)$ yields a stable series in powers
of $1/p$.  To within $O(1/p^7 )$, the expansion 
for (\ref{eq:ARp}) is:
\begin{eqnarray}
A(R,p) & = &
(p+1)^2 -2\,R-\frac{(4\,R+1)}{4 \, p}+\frac{(2\,R+1)}{4 \, p^2}
-\frac{(16\,R^2+40\,R+23)}{64 \, p^3} \nonumber \\
& + & \frac {(32\,R^2+68\,R+41)}{64 \, p^4}
-\frac{(64\,R^3+576\,R^{2}+1108\,R + 681)}{512 \, p^5} \nonumber \\
& + & \frac{(256\,R^3+1432\,R^2 + 2566\,R + 1593)}{512 \, p^6} \label{eq:ARpser} \\
& - & 
\frac{(1280\,R^4 +28160\,R^3+123680\,R^2+ 210448\,R+131707)}{16384 \, p^7}
\nonumber\\
& + & \frac{(8192\,R^4+95040\,R^3+358368\,R^2 + 587512\,R+371061)}{ 16384 \, p^8}
\nonumber\\
&  - & \frac{(7168\,R^5+313600 \,R^4+ 2607232\,R^3+8854496\,R^2 +14149364\,R+9039151)}{ 131072 \, p^9}
\nonumber\\
& + & \frac{(65536\,R^5 +1366016\,R^4 +9200576\,R^3+29011472\,R^2 45621790\,R
+29559559)}{131072 p^{10}} \, . \nonumber
\end{eqnarray}
The coefficients up to $O(p^{-6})$ have been previously 
published\cite{taylorc}
but our computer algebra programs allow us to go much further.

\subsection{Other Bases - Algebraic Vindication}
Although our previous results are apparently independent
of the size of the chosen basis, we must consider 
other bases.  For the $\eta$ coordinate,
we consider the Baber-Hass\'{e} and
the Wilson bases\cite{moniquea} which are
described as follows. \\

\noindent
\underline{Baber-Hass\'{e}:} 
\begin{equation}
M(\eta , \phi ) = e^{i m \phi } e^{-q \eta } 
~ \sum_{\ell=m} a_{\ell } ~ P_{\ell}^m (\eta )
\end{equation}
The recurrence relation is given by:
\begin{equation}
\frac{(\ell + m +1) }{(2 \ell +3) } 
[ 2 p (\ell + 1) + \Rm_{1} ] a_{\ell+1} +
\alpha_1 (k) a_{\ell} +\frac{(\ell -m)}{(2 \ell -1)} (\Rm_{1} - 2 p \ell ) a_{\ell-1} = 0
\end{equation}
where for $m=0$:
\begin{eqnarray}
\alpha_1 (k) & = & A~-~p^2 ~+~ \ell (\ell +1 ) \nonumber \\
a_{-1} & = & 0 ~. \nonumber
\end{eqnarray}

\noindent
\underline{Wilson}: 
\begin{equation}
M(\eta , \phi ) = e^{i m \phi } e^{q \eta } 
(1 - \eta )^{m/2} ~ \sum_{k=0} (-1)^k ~ c_k ~ (1- \eta)^k
\end{equation}
The recurrence relation is:
\begin{equation}
2 (k+1)~ (k+ m + 1) ~ c_{k+1} + \sigma_1 (k) ~c_k +2 [ \Rm_{1} ~+~ p ~(k+m) ~c_{k-1} = 0
\end{equation}
where
\begin{eqnarray}
\sigma_1 (k) & = & A~-~p^2 ~+~ (m+1)~ (m+2 p) + ~k~ (k+2~m+4 ~p ~+~1) \nonumber \\
c_{-1} & = & 0 ~. \nonumber
\end{eqnarray}
Both of these bases have been implemented into the Maple system. If
we consider $\ell=0$, the coefficient $a_{N}$ of Baber-Hass\'{e} basis 
is of order $O(1/p^N )$ and the coefficient $c_{N}$ of Wilson basis 
is of order $O(p^0 )$. However, if we inject our series solution 
for $A(p^2 )$ into the series coefficients of both bases, we find
that both $a_N$ and $c_N$ are {\em formally} zero to within order $O(p^N )$.
This can be seen through a number of computer algebra demonstrations.
Thus our series solution for $A(p^2 )$ also {\em formally} satisfies 
the recurrence relations of these other bases, {\em order by order} in $p$.

For the $\xi$ coordinate, apart from the used Hylleraas basis,
there is also the  Jaff\'{e} basis.\\

\noindent
\underline{Jaff\'{e}}: 
\begin{equation}
\Lambda (\xi ) =  (\xi^2-1)^{m/2} (\xi +1)^{-m-1+\Rm_2 / 2p}
~ e^{-p \xi}
\sum_{k=0} D_k \left( \frac{\xi-1}{\xi+1} \right)^k
\end{equation}
The recurrence relation is:
\begin{equation}
(k + 1)~ (k+ m+1 ) ~  D_{k+1} + \gamma_1 (k) ~D_k  
  +   (k - \frac{\Rm_2}{2p} ) (k + m - \frac{\Rm_2 }{2p} ) D_{k-1} 
\end{equation}
where
\begin{eqnarray}
\gamma_1 (k) & = & A~-~p^2 + \Rm_{2} ~-~ (m+1)~(2 p +1 - \frac{\Rm_{2}}{2 p})
\nonumber \\
& - & 2 k~ (k+ m +2 p +1 -\frac{\Rm_2}{ 2p}) \nonumber \\
D_{-1} & = & 0 ~. \nonumber
\end{eqnarray}
Similarly, it can be algebraically demonstrated that \eg{}
for $\j = n -\ell - 1 = 0$,
the $1/p$ series expansion of $A(R,p)$ 
formally satisfies the coefficients of the Jaff\'{e} basis 
for negative powers of $p$ just as they satisfy the Hylleraas basis.
This demonstration allows us to consider
another basis of importance for the $\eta$ coordinate, namely
the Power basis:\\

\noindent
\underline{Power}: 
\begin{equation}
M(\eta , \phi ) = e^{i m \phi } e^{-q (1+ \eta ) }
(1 - \eta )^{m/2} ~ \sum_{k=0} ~ d_k ~
{\bf M } (-(k + \delta k ), m+1 , 2 p (1+ \eta) )
\end{equation}
$M$ is the confluent hyergeometric function. The recurrence relation is:
\begin{equation}
(k + \delta k +1)~ (k+ \delta k + 1 - \frac{\Rm_{1}}{2 p} ) ~
d_{k+1} + \chi_1 (k) ~d_k ~ \quad \quad ~
\end{equation}
\[
~ \quad \quad \quad \quad + (k + \delta k + m )
(k+ \delta k + m - \frac{\Rm_{1}}{2 p} ) ~ d_{k-1} 
\]
where
\begin{eqnarray}
\chi_1 (k) & = & A~-~p^2 - \Rm_{1} ~+~ (m+1)~(2 p -1 + \frac{\Rm_{1}}{2 p}) 
\nonumber \\
& - & 2 (k+ \delta k )(k + \delta k +m+1 -2 p - \frac{\Rm_{1}}{2 p}) \nonumber \\
d_{-1} & = & 0 ~, \nonumber
\end{eqnarray}
 and $\delta k$ is an exponentially vanishing term in $R$ and consequently
we do not make the same demonstration as
for the Wilson and Baber-Hass\'{e} bases.  However, when we let
$R \rightarrow \infty$ then $\delta k \rightarrow 0$ and we can make
a similar demonstration as for the Jaff\'{e} basis using the $1/p$
expansion of $A(R,p)$.

Granted, we have not proven this for {\em all} bases. 
Nonetheless, we emphasize that
\eg{} the Wilson basis is very different from the Baber-Hass\'{e}
basis or the Power Basis and the basis of spherical harmonics we 
used as a starting point for this analysis.  Moreover, the Hylleraas basis
is also very different from the Jaff\'{e} basis.
These demonstrations strongly
suggest basis independent results for $A(p^2 )$ and $A(R, p)$.

This analysis herein exploits the fundamental
theorem of algebra \ie{} that if one knows all the $N$ roots of
a given polynomial say $P_N (x)$, the latter is completely defined within
a scaling factor namely the coefficient of its highest power in $x$.
The three-term recurrence relations of eqs.~(\ref{eq:dFe}), (\ref{eq:dFo}) 
and (\ref{eq:dY}) have a linear dependence on $A$ for the term in $d_k$ 
but {\em no} dependence of $A$ for the third term in $d_{k-1}$.
Thus, $\dFe{i}$, $\dFo{i}$ and $\dY{i}$ are $i^{th}$ degree polynomials 
in $A$ regardless of whether or not the third term in $d_{k-1}$ is neglected.
This allows us to completely account and identify the 
the eigenparameters of the matrices $\F$ and $\Y$ for every discrete state.

\subsection{Mathematical Classification of Solutions}
So far, we have identified the functions implied by
the determinants $\dF{i}$ and $\dY{i}$, namely
$A(p^2 )$ and $A(R,p)$ respectively for {\em all} discrete states
where $m=0$ for the homonuclear case. In view of
previous and recent work on the $D \rightarrow 1$ version
of H$_2^+$ and the findings in this work concerning
the $D=3$ version of H$_2^+$, we are now equipped with
the means to make the following analytical comparison.
Here, we can put the $D \rightarrow 1$ and the $D=3$ versions
of H$_2^+$ on the same ``canonical'' footing:
\begin{description}
\item[D $\rightarrow$ 1:]
To reiterate the results of section \ref{sec:prel},
the energy eigenvalue is governed by an equation of the form:
\begin{equation}
\exp (-2 ~ R ~ d) = P_2 (d)_{\{P_N (d)\}}  \quad \mbox{where} \quad E = -\frac{d^2}{2}
\label{eq:D1}
\end{equation}
When the second order polynomial $P_2 (d)$ factors into a product of
first order polynomials, both sides of eq.~(\ref{eq:D1}) factors and
the solution for $d$ is a (standard) Lambert $W$ 
function\cite{lamberta,lambertb}. 
When it does not factor, the solution is a generalization of the $W$ function 
reported in the work of \cite{mann}.
When the right side is a polynomial, the  
solution is a {\em generalized} Lambert $W$ function\cite{martinez}. 
The subscript $P_N (d)$ reminds us that our generalization for
the $W$ function can accommodate a polynomial with rational
coefficients of arbitrary degree on the right side of eq.~(\ref{eq:D1}).

The exponential term on the left side is a reflection of the
fact that outside the Dirac delta function wells, the basis of
the particle is a combination of free particle solutions which
required matching at the Dirac delta function peaks.
\item[D=3:]
To summarize the results of the past few sections, the eigenparameter
$p$, which plays an analogous role to the parameter $d$ of the
$D \rightarrow 1$ version of H$_2^+$ is determined from the
equation:
\begin{equation}
A(R,p) = A(p^2 )_{ \{P_N (p) \}} \quad \mbox{where} \quad E = -2~\frac{p^2}{R^2}
\label{eq:D3}
\end{equation}
The subscript $P_N (p)$ reminds us that we have a Taylor series
for $A(p^2) $ with rational coefficients which exactly matches the
generalized right-hand side form of eq.~(\ref{eq:D1}).
However, the left-side of (\ref{eq:D3}) looks very different
than the left side of (\ref{eq:D1}); it is the function 
implied by $\dY{i}$ and is associated
with the separated quantum number $\j = n- \ell -1$.  
Nonetheless, like $\exp (-2 R d)$, this function is well-defined
asymptotically for large $R$.
The right side
of eq.~(\ref{eq:D3}) is implied from $\dF{i}$ for even or odd $\ell$ which
is a united atom quantum number.  

So far, the functions $A(R,p)$ and $A(p^2 ) $ appear in the literature as 
expansions in terms of $p^{-k}$ and $p^k$ respectively, restricted to 
$k=6$ and for specific cases of large and small 
values of $R$\cite{taylora,taylorb,taylorc}.
We can obtain series representations of both to a much
greater extent in view of our computer algebra implementations.
We have also seen that $A(p^2 )$ 
can be represented as an infinite series in $x$ where $x=p^2$ and
is consequently polynomial-like.  

Note that if $m \ne 0$, the governing equation has the same form as
eq.~(\ref{eq:D3}) but the left side is more complicated and
more difficult to get, as the determinant $\dY{i}$ is no longer
governed by a simple recurrence relation.  However, in principle,
eq.~(\ref{eq:D3}) governs the entire homonuclear case.
\end{description}

Mathematically, in both cases, the right side of the governing
equation is expressed in terms of only one of the eigenparameters 
whereas the left side requires the parameter and the value
of the internuclear distance $R$ which is determined on input.

We therefore come to the conclusion that the eigenparameter $p$, like
its $D \rightarrow 1$ counterpart $d$, is also determined by a
special function which is an {\em implicit} 
function, an even greater generalization of the Lambert W function.
So far, the functions $A(R,p)$ and $A(p^2 )$ do not appear in the literature.
However, we can obtain series representations of both to the
extent of getting reliable numbers, as demonstrated by our tables
of values. 

On the subject of implicit functions or implicit equations, 
these are seen in a number of specific contexts:
\begin{description}
\item[Retardation Effects:]
Equations of form $e^{-c~\lambda} = P_2 (\lambda)$ and more generally
$e^{-c \lambda } = P_2 (\lambda ) / Q_1 (\lambda )$ express the solutions of
a huge class of {\it delayed} differential equations\cite[eq.(3)]{sueann}.
\item[Bondi's K-calculus:]
It is well known in the area of special relativity that 
the Lorentz transformation can be derived from the 
theory of implicit
functions with minimal assumptions of continuity\cite{bondi}.  
Here one seeks the function $f(t)$ satisfying $f(f(t)) = k^2 t$
and the requirements
that it be monotone increasing and continuous. The unique solution 
is:
\[
f(f(t)) =  k^2 t  \quad \mbox{where} \quad v/c = k^2 - 1/k^2
\quad \rightarrow \quad f(t) = k~t
\]
\item[GRT/QFT:]
As we mentioned before, the Lambert W function and its
generalization appear in General Relativity
as solutions to respectively the two-body and three-body
linear $(1+1)$ gravity problems via dilaton theory\cite{mann,mann2}.  
\end{description}
Implicit functions often appear in problems with retardation
effects, relativistic or otherwise. 
Thus, with some reservations, we associate with this mathematical
category a tentative ``physical picture'':

Although the hydrogen molecular ion H$_2^{+}$ in the context
of the Schr\"{o}dinger wave equation is {\em not} a relativistic
formulation, the eigensolutions we obtain nonetheless
suggests something akin to a {\em retardation} or {\em delay} 
effect.  This is not the case for a one center problem
like the hydrogen atom but this characteristic appears 
for a two-center problem.  However, this statement must be tempered 
with the fact that \eg{} the Lambert W function also appears in 
many other types of problems with no relationship to 
retardation effects.


\addtocontents{toc}{\vspace{0.3cm}}
\section{Summary/Conclusions}

Through experimental Mathematics using computer algebra
as a tool, we have identified the mathematical {\em structures} 
governing the energy spectrum of the hydrogen molecular ion
H$_2^+$ for the two-center one-electron problem.  

In the present work, we started with a 
particular choice of basis and expressed the determinantal conditions by 
which the eigenparameters $p$ and $A$ are obtained.  From one of the
two determinants, we inverted the problem to obtain a 
series representation of the separation constant $A(p^2 )$
associated with the united atom quantum number $\ell$.
We applied a similar approach to obtain $A(R,p)$ from the remaining
determinant and associated with the separated atom 
quantum number $\j = n - \ell -1$.
where $n$ is a united quantum number.  
We then demonstrated that the results were independent of the size
and even the choice of basis. 

The eigenparameter $p$ for which $E=-2p^2/R^2$
is obtained by matching $A(p^2 )=A(R,p)$ and found to be the
solution of an implicit function, with features similar
to that of the Lambert W function {\em and} its recent
generalizations\cite{martinez}.  
This allowed us to mathematically categorize the 
eigenvalues (or rather make us realize what they are not) 
and even to associate a tentative ``physical picture'' to the
solutions.  While we made no pretense at rigor, the solutions were 
nonetheless vindicated numerically and by algebraic demonstrations 
with computer algebra.

The results express analytical solutions for the ground state
and the countable infinity of discrete 
states of H$_2^{+}$ for the homonuclear case when the
magnetic quantum number $m=0$.  From the
discussion below eq.~(\ref{eq:D3}), we anticipate that 
the eigensolutions for $m \ne 0$ for the homonuclear case to
be qualitatively similar though admittedly this remains
to be proven.  We emphasize that although the basis and 
approach used here were ideal for $m=0$ and the homonuclear case.
the computer algebra methods
shown are directly applicated to the heteronuclear case with
$m \ne 0$.  
For $m \ne 0$, one should work directly from the recurrence
relations of the chosen basis now that we understand how these
basis coefficients behave with better and better accuracy for
the series expansions of $A( p^2 )$ in $p$ and the asymptotic series
expansions for $A(R,p)$ in $1/p$.

However, we make no pronouncements concerning the nature
or mathematical category of the solutions for the
heteronuclear
case or when the nuclei are allowed to move.
We note that for $m=0$, the matrix for $\dY{i}$ 
remains tridiagonal while the band matrix for $\dF{i}$ 
is pentadiagonal and consequently governed by nested recurrence 
relations\cite{moniqueb} suggesting that the analysis
shown herein is possible.

A number of issues arise from this result.  In a sense,
the result is both overdue and premature.  It is overdue
because of our present capacity to find solutions to
fair-sized molecules using computational chemistry. On
the other hand, it is premature.  The functions we found
$A(p^2 )$ and $A(R,p)$ do not seem to resemble anything we
have seen in the literature.  The apparent singularities
or ``resonances'' at $2 p^2 = 3 \, (4 \, j + 3) ( 4 \, j + 7 ) $ 
for $A(p^2 )$ and $R = (j+2) \, p +2~p^2$ for $A(R, p)$
for $j=0,1,2,\ldots $ do not constitute
a problem since the eigenparameters $p$ and $A$ for a given
$R$ are never found on these resonances.  Once a value
of $R$ is injected into $A(p^2 ) =A(R,p)$, solving for $p$
numerically did not create any problems in the test cases
examined so far.  At any rate, the tables shown herein 
merely illustrate the convergence properties of the functions
$A(p^2 )$ and $A(R,p)$ we have identified:
solving the coupled set of 
polynomials $\dF{i}$ and $\dY{i}$ for $p$ and $A$ at a
given distance $R$ involves no resonances
and is still the most useful method from a computational 
point of view. In principle, the latter can go further than
any FORTRAN program.

We have
ordered series representations to relatively high
order of both of these functions $A(p^2 )$ and $A(R,p)$ and 
we can generate reliable numbers for a number of discrete quantum 
states.  We have also demonstrated that we could
use these series beyond their radius of convergence using
techniques for handling divergent series.

From here, one could explore and seek alternate representations
of these functions with better convergence properties especially at low
$R$ for $A(R,p)$ and large $R$ for $A(p^2 )$ but the results from the 
Sidi transformations are already very promising.  At any rate, 
the hydrogen molecular
ion for clamped nuclei can be entirely contained within simple
computer algebra sessions, not much more complicated than those
of the hydrogen atom\footnote{Maple CAS programs are available upon request.}.

The exploratory and roundabout way 
by which we found our solutions, suggests there
is something missing in the mathematical physics or the methods for
obtaining the eigenvalues of the Schr\"{o}dinger wave equation.  
There is hardly any existing ``technology'' for solving 
quantum chemistry problems involving implicit functions.
Our use of a basis is certainly valid to demonstrate or prove a
result. Furthermore, the convergence of the bases used here has been
confirmed by determining
the asymptotic behavior of the expansion
coefficients of the wavefunctions for the
various basis sets considered\cite{taylorc}.
Nonetheless, a more direct way of generating the functions
of $A(p^2 )$ and $A(R,p)$ would be instructive.

\addtocontents{toc}{\vspace{0.3cm}}
\addcontentsline{toc}{section}{\numberline {Acknowledgments} \hspace*{2cm} }
\section*{Acknowledgments}
One of us (T.C.S.) would like to thank 
Professor Arne L\"{u}chow
of the Institut f\"{u}r Physikalishe Chemie, RWTH Aachen
and Professor Georg Jansen of the Institut f\"{u}r Organische Chemie
of the University of Essen for their wonderful hospitality and
support for allowing this work to be possible.
Special thanks go to Marc Rybowicz of the University of
Limoges for helpful discussions concerning the recurrence
relations and Bruno Salvy for his ``gfun'' package and
discussions with Philippe Flajolet on asymptotic expansions.
We would also like to thank Dirk Andrae
of the Theoretical Chemistry
group at the University of Bielefeld (Faculty of Chemistry), 
and Greg Fee of the Center of Experimental 
and Constructive Mathematics (CECM) of Simon Fraser University,
for helpful information.


\addtocontents{toc}{\vspace{0.3cm}}
\addcontentsline{toc}{section}{\numberline {References} \hspace*{1cm}}

\newpage
\newpage

\begin{figure}
\begin{center}
\underline{\large \bf Energy vs. R for H$_2^{+}$ (a.u.) } 
\[
\mbox{\color{red} \underline{~~~~~~~~} \color{black}  {\bf D} } \rightarrow \infty 
\quad  
\mbox{\color{blue} \underline{~~~~~~~~} \color{black} {\bf D} } \rightarrow 1 
\quad
\mbox{\color{green} \underline{~~~~~~~~} \color{black} {\bf D} = 3}
\]
\end{center}
\centerline{\psfig{figure=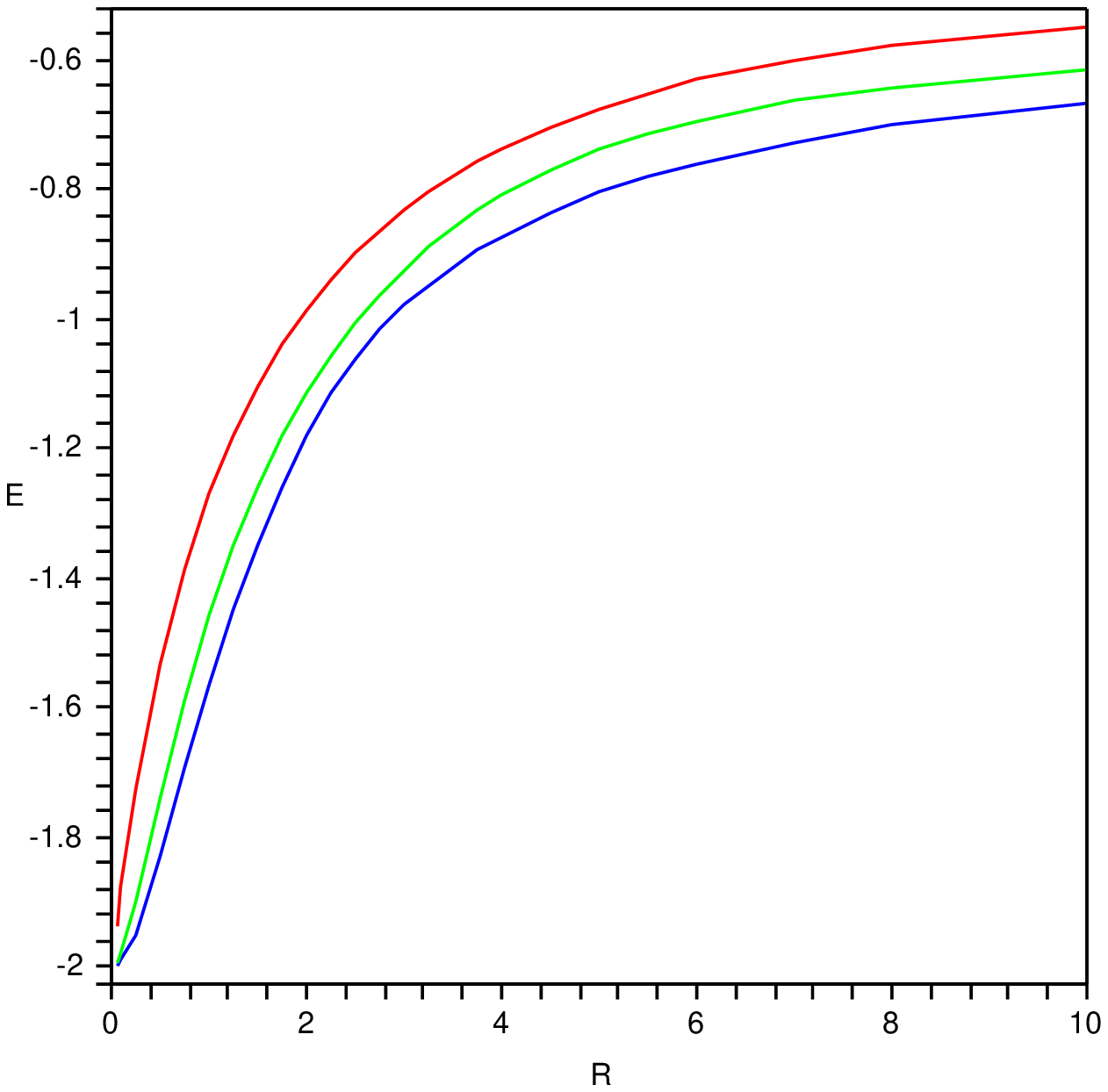,height=6.1in,width=7.0in}}
\[
\mbox{\color{green} \bf E$_{\bf 3}$ \color{black}} (R)
\approx \frac{1}{3} 
\mbox{\color{blue} \bf E$_{\bf 1}$ \color{black}} 
({\textstyle \frac{R}{3}})~+~
\frac{2}{3} 
\mbox{\color{red} \bf E$_{\bf \infty}$ \color{black}}
({\textstyle \frac{2 \, R}{3}})
\]
\underline{Reference:} 
Lopez-Cabrera, Tan and Loeser, {\em J. Phys. Chem.} {\bf 97}, 2467-2478 (1993).
\end{figure}

\end{document}